\documentstyle[psbox,amssymb]{WORKSHOP}

\newcommand{\atel}{ATel}
\newcommand{\mnras}{MNRAS}
\newcommand{\apss}{Ap\&SS}

\begin{document}

\title{
\emph{Swift} observations of black hole candidate XTE J1752-223 during outburst
}

% AUTHOR(S) 
\author{
P.A.~Curran$^1$, 
P.~Casella$^2$,
T.J.~Maccarone$^2$ 
P.A.~Evans$^{3}$, 
W. Landsman$^{4}$, \\ 
H.A.~Krimm$^{5,6}$,  
C.~Brocksopp$^{7}$,
and M.~Still$^{8}$
\\[12pt]  % TO BE SPACED WITH ONE LINE
%
% INSTITUTES OF AUTHORS
$^1$ AIM, CEA/DSM - CNRS, Irfu/SAP, Centre de Saclay, Bat. 709, FR-91191 Gif-sur-Yvette Cedex, France \\
$^2$ School of Physics and Astronomy, University of Southampton, Southampton, Hampshire, SO17\,1BJ, UK \\
$^3$Department of Physics and Astronomy, University of Leicester, University Road, Leicester LE1\,7RH, UK \\
$^4$Adnet Systems, NASA/Goddard Space Flight Center, Code 667, Greenbelt MD 20771, USA \\
$^5$NASA/Goddard Space Flight Center, Astrophysics Science Division, Code 661, Greenbelt, MD 20771, USA  \\
$^6$Universities Space Research Association, Columbia, MD 21044, USA \\
$^7$Mullard Space Science Lab, University College London, Holmbury St Mary, Dorking, Surrey RH5\,6NT, UK\\
$^8$NASA Ames Research Center, Moffett Field, CA 94035, USA \\
%
% please put the first author's initial and e-mail address below
{\it E-mail(PAC): peter.curran@cea.fr} 
}

\abst{
Here we summarise the \emph{Swift} broadband observations of the recently
discovered X-ray transient and black hole candidate, XTE\,J1752-223,
obtained over the period of outburst from October 2009 to June
2010. 
We offer a  phenomenological treatment of the spectra as an
indication of the canonical spectral state of the source during
different periods of the outburst. We find that the high energy
hardness-intensity diagrams over two separate bands follows the
canonical behavior, confirming the spectral states. 
From \emph{Swift}-UVOT data we confirm the presence of an
optical counterpart which displays variability correlated, in the soft
state, to the X-ray emission observed by \emph{Swift}-XRT. The optical
counterpart also displays hysteretical behaviour between the states
not normally observed in the optical bands, suggesting a possible
contribution from a synchrotron emitting jet to the optical emission
in the rising hard state. 
Our XRT timing analysis shows that in the hard state there is significant variability
below 10\,Hz  which is more pronounced at low energies, while during
the soft state the level  of variability is consistent with being
minimal.These properties of XTE\,J1752-223 support its candidacy as a
black hole in the Galactic centre region.
}

\kword{X-rays: binaries --- X-rays: bursts --- Individual: XTE J1752-223}

\maketitle
\thispagestyle{empty}

\section{Introduction}\label{section:introduction}

Low mass X-ray binaries are for the majority of the time in a state of quiescensce with faint or non-detected X-ray emission, though optical or near-infrared (nIR) counterparts may be visible due to emission from the donor star, or possibly the jet, hot spot, or outer accretion disk. They are often only discovered when they enter an active state of outburst when -- powered by an increased level of accretion onto the central, compact object (black hole or neutron star) -- there is a dramatic increase of the X-ray, optical/nIR and radio flux. 
During these outbursts the systems have been observed to go through a number of high energy spectral states before returning to a quiescent state, usually on times scales of weeks, months or even longer. 
These states are a generally low intensity, power-law dominated, {\it hard } state followed by a usually, higher intensity, {\it thermal-dominant}, {\it soft} state which decreases in flux, via a late hard state, over time. Additionally, the hard states are  associated with aperiodic variability of the light curve not present in the soft state see (McClintock \& Remillard 2006 for a fuller description of the various possible states).

XTE\,J1752-223, a new X-ray transient and black hole candidate in the Galactic
center region, was detected on 2009-10-23 at 19:55 UT
(MJD 55128.33) by RXTE (Markwardt et al. 2009a). The high energy, variable emission of the
source was confirmed in the following days by \emph{Swift}-XRT \&
BAT (Markwardt et al. 2009b)
and RXTE (Remillard et al. 2009, Shaposhnikov et al. 2009)
as well as by MAXI/GSC (Nakahira et al. 2009) and
\emph{Fermi}/GBM (Wilson-Hodge et al. 2009). 
An optical and nIR counterpart was proposed (Torres et al. 2009a, 2009b)
and a radio source coincident with the X-ray position,
later confirmed as the jet (Yang et al. 2010), was detected (Brocksopp et al. 2009). The source was observed to
have undergone a state transition in mid-January 2010 (MJD $\sim 55210$),
from a spectrally hard to a spectrally soft X-ray state (Homan et al. 2010) . This was
identified by a decoupling of high energy ($\gtrsim 4$\,keV) and low
energy ($\lesssim 4$\,keV) MAXI light curves which had previously traced each other, and by
the dominance of a thermal component in spectra. The source was
observed to have reverted to a hard state at the end of March 2010
(MJD $\sim 55280$; Mu{\~n}oz-Darias et al. 2010), after the low energy light curve decreased to trace 
once more the high energy light curve; the thermal component was no longer dominant.
Here we summarise our paper (Curran et al. 2010) detailing the \emph{Swift} -- Burst Alert Telescope (BAT), X-ray Telescope (XRT) and Ultraviolet/Optical Telescope (UVOT) -- monitoring
observations of XTE\,J1752-223, obtained over the period of the
outburst. Based on these data, we identify the periods of the
various states (McClintock \& Remillard 2006) and compare the behavior of the major photometric,
spectral and timing properties during these states to those expected
from black hole X-ray binaries.

\section{Defining the states}\label{sec:states}

\begin{figure*} 
  \centering 
  %\psbox[xsize=8cm]{fig.spectra_00.ps} 
  \psbox[xsize=16cm]{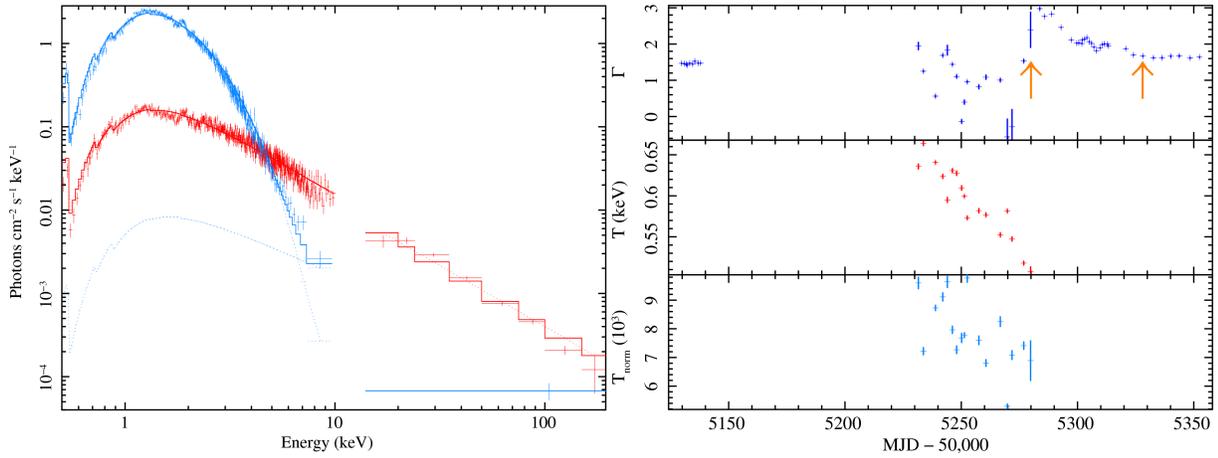} 
\caption{
    {\it Left}: 
    Example XRT-BAT spectra in the power-law dominated, hard state (red; MJD 55131) and in the soft state (blue; MJD 55248) which displays a strong thermal component. 
    {\it Right}: 
    Best fit parameters of the joint XRT-BAT spectral fits: power law photon index, $\Gamma$,  thermal component temperature, $T$, and thermal component normalization, $T_{{\rm norm}}$, in units of $10^{3}$. 
    Note that the photon indices, in the period where there is a thermal component, are not reliable. 
    The absence of data from MJD 55139 to 55231 is due to the various instruments becoming sun-constrained. The arrows signify the dates between which the source is transitioning from the soft state to the late, hard state.
}
 \label{fig:spectra} 
\end{figure*}

The states are defined by the XRT-BAT spectral analysis (Figure \ref{fig:spectra}) which shows that in the initial
hard state (MJD 55131 - 55138), the spectra are dominated by a single, hard power law
component of photon index, $\Gamma=1.46 \pm 0.03$, absorbed by a column density of $N_{{\rm H}} = 0.513 \pm 0.003 \times 10^{22}$\,cm$^{-2}$. 
In the soft state (MJD 55233 - 55280) there is a
significant additional contribution from a thermal component at 
$\sim0.6$\,keV, where the absorption is fixed at the previous value.
In the intermediate state (MJD 55283 - 55328), a thermal component in no longer
supported though the power law is still decreasing until it reaches its
final, hard state (MJD 55329 - 55352) value of $\Gamma=1.64 \pm 0.02$ 
where the column density is again fixed to that of the initial hard state. 
This steeper value at late times,
like the value of hardness ratio (HR; section \ref{sec:hid}), shows that the source has not returned to
it's initial state or that the late hard state has different 
spectral properties, such as column density, than the initial hard state.
Leaving column density free in the late spectral fits leads to a shallower photon index, 
consistent with the original, of 
$\Gamma=1.53 \pm 0.02$ absorbed by a column density of $N_{{\rm H}} = 0.43 \pm 0.01 \times 10^{22}$\,cm$^{-2}$. More detailed spectral fitting with better constrained photon indices are required to differentiate whether the column density is variable or the photon index is different at late times.

\begin{figure*} 
  \centering 
    %\psbox[xsize=8cm]{fig.xray_3.ps}
    %\psbox[xsize=8cm]{fig.powspec.ps}
  \psbox[xsize=16cm]{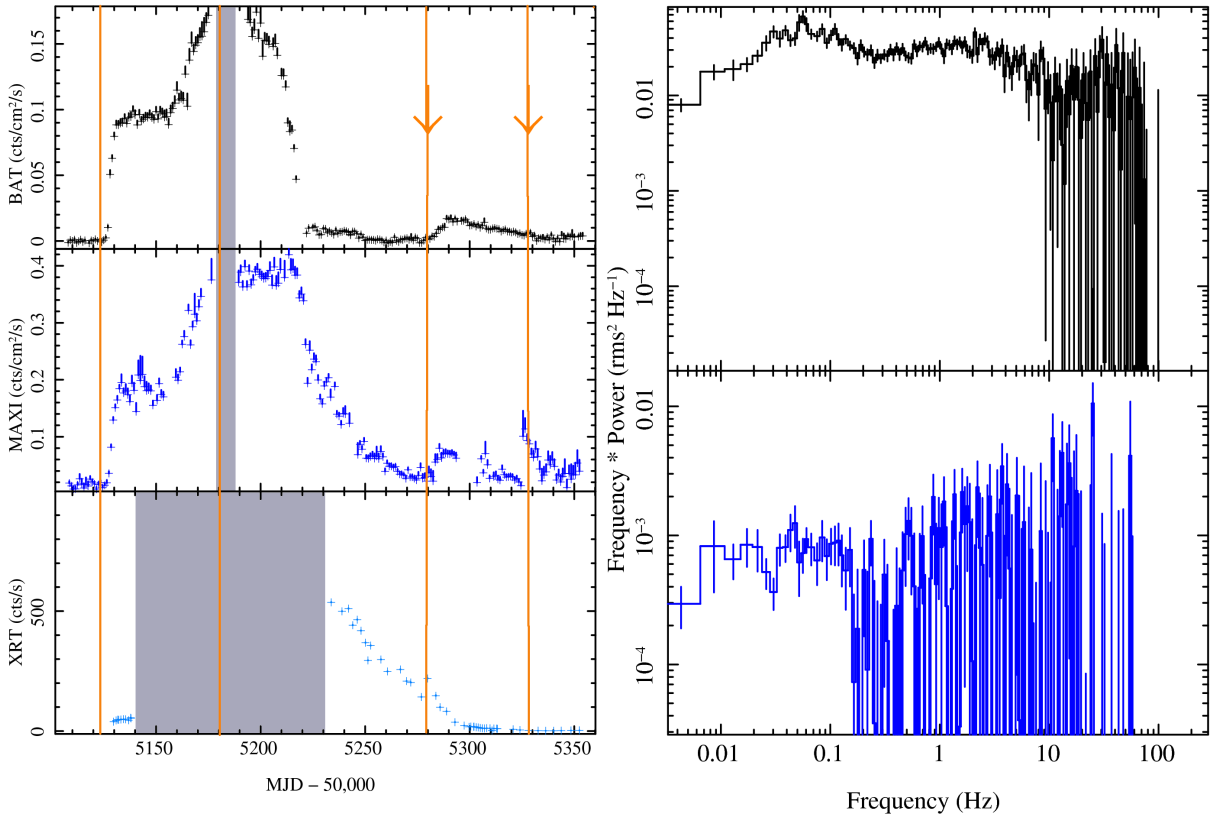}
   \caption{
    {\it Left}: 
    High energy (\emph{Swift}-BAT [15-150\,keV], MAXI
    [4-10\,keV], \emph{Swift}-XRT [0.3-10\,keV]) light curves during
    the period of outburst. 
    The absence of data around MJD
    $\sim$55180 (shaded area) is due to the various instruments becoming
    sun-constrained. The orange lines signify the dates of the state transitions: 
    quiescence -- hard state -- soft state -- intermediate state -- hard state.
    {\it Right}: 
    The average power density spectrum (PDS) for XRT light curve in
    the initial hard  state ({\it upper}; RMS $\sim 54$\%) exhibits the
    aperiodic variability of the light curve not present in the
    average soft state PDS ({\it lower}; RMS $<12$\%).
  }
  \label{fig:xray}  
\end{figure*}

In the high energy \emph{Swift}-BAT [15-150 keV], MAXI [4-10keV] and \emph{Swift}-XRT
[0.3-10 keV] light curves (Figure \ref{fig:xray}) during the period of outburst, the
absence of data at MJD 55139-55233 (shaded area) is due to the various
instruments becoming sun-constrained. The orange lines signify the
dates of the state transitions: quiescence -- hard state -- soft state --
intermediate state -- hard state. It is clear from these that there is
an excess of soft/low energy photons in the soft state as exhibited by
the decoupling of high energy and low energy light curves which had
previously traced each	other. Unfortunately, due to the sun
constraint, the intermediate state between the initial hard state and
the soft state was unobserved.

\section{Optical counterpart \& hysteresis}

\begin{figure*} 
  \centering   
  %\psbox[xsize=8cm]{fig.UVOT-XRT.ps}
  %  \psbox[xsize=8cm]{fig.lc_opt.ps} 
  \psbox[xsize=16cm]{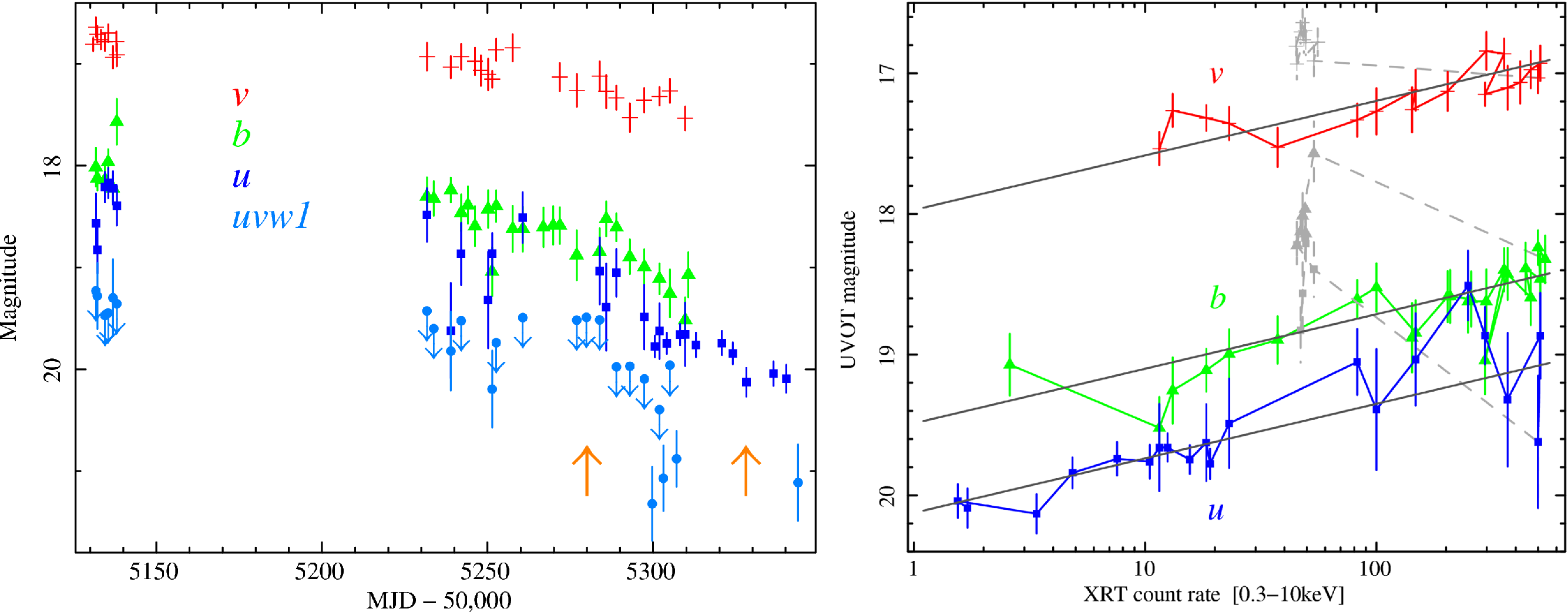}
  \caption{
    {\it Left}: 
    UVOT  $v$ (crosses), $b$ (triangles), $u$ (squares) and
    $uw1$ (circles with $3\sigma$ limits) band light curves for the source
    show variability; dimming by 1 magnitude or more from the
    start to the end of observations. The up-pointing arrows signify
    the dates between which the source is transitioning from the soft
    state to the late, hard state. 
    {\it Right}:
    UVOT magnitude in three filters versus XRT count rate. Data points in grey
    indicate observations before MJD $\sim 55230$, i.e., the initial
    hard state. The solid black lines represent the simultaneous power
    law fit to the data after MJD $55230$.
  }
  \label{fig:uvot-xrt} 
\end{figure*}

UVOT $v$, $b$, $u$ and $uvw1$ band light curves (Figure \ref{fig:uvot-xrt}) for
the proposed optical counterpart allowed us to confirm the association
with XTE\,J1752-223 due to their variability. 
The accurate position of the counterpart was derived from a deep 
(1853\,s) $v$ band image as 17:52:15.08  $-$22:20:32.9 (J2000; 0.31\,arcsecond $1\sigma$ error). 
Given the unknown Galactic extinction to the source and the quality of the data, we
cannot examine the true colours or spectral shape of the optical
source, neither can we confirm any possible spectral changes over the
outburst. Assuming a power law spectrum, the spectral index could be
anywhere between a rising value of 2.0 and a decreasing value of 5.3.

The correlation between the UVOT magnitudes in three bands and the 
XRT count rate during the soft state (Figure \ref{fig:uvot-xrt}) further
confirm the	association	between	the proposed counterpart and
XTE\,J1752-223. 
Such a correlation is more usually associated with the hard state 
(Russell et al. 2006) but we cannot say if the early, hard 
state data follows another correlation due to the very limited range 
of magnitudes and X-ray count rates over this period.
Data points in grey indicate observations during the
initial hard state and clearly exhibit a hysteresis with the later
soft state data at a similar count rate.
 This is similar to the hysteretical behaviour observed in the nIR for a number of transients (Russell et al. 2007), most notably XTE\,J1550-564,   where the additional hard state emission is attributed to optically thin synchrotron emission from a jet, which would be quenched in the soft state, and weak at low X-ray luminosities, leading to the hysteresis. 
Though this hysteresis is not normally observed in the optical bands, the optical data presented here is in good agreement with the synchrotron emitting jet making a significant contribution to the optical emission in the rising hard state.

\section{X-ray variability}

The average power density spectrum (PDS; Figure \ref{fig:xray}) for XRT light curve in the
hard state exhibits an aperiodic variability of the light curve (RMS
$\sim54$\%) at frequencies up to 10\,Hz, though no QPOs are
detected. The average soft state PDS exhibits only minimum variability
and only at frequencies below 0.2\,Hz (RMS $<12$\%). These PDS help
confirm the identified states as they exhibit the same properties as
observed in other black hole X-ray binaries.
We also find that, in the initial hard state, there is an excess of
variability at low energies: the low energy (0.3-1.5\,keV) light curves
having 20\% higher RMS than the high energy (1.5-10\,keV) light
curves. This may suggest a contribution from intrinsic disk
variability which is not obvious from the spectral fitting
(section \ref{sec:states}). 
More detailed spectral fitting and better constrained column
density are required to test this possibility.

\section{Hardness-Intensity Diagrams}\label{sec:hid}

\begin{figure*} 
  \centering  
  \psbox[xsize=16cm]{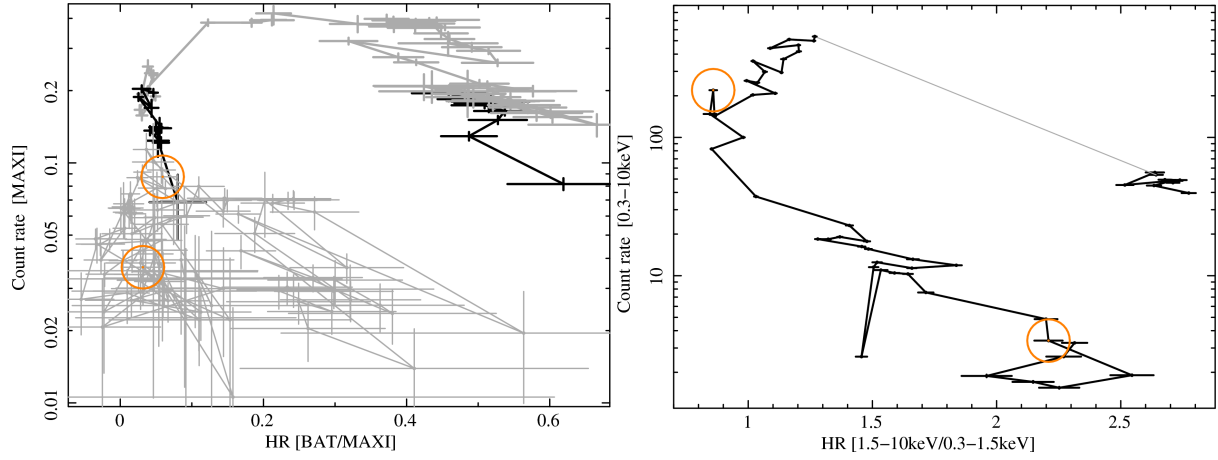}
  %\psbox[xsize=8cm]{fig.HID-XRT.ps}
  %\psbox[xsize=8cm]{fig.HID-BAT_MAXI.ps}
\caption{
  Hardness-Intensity diagrams (HID) for both BAT/MAXI ({\it left}) and
  XRT ({\it right}). The grey lines, at high count rates in 
  both plots show the time during which the XRT was unable to
  observe (MJD 55139-55233),  while the circles show the start and end
  of the transition from the soft state.  Note
  that at low count rates ($\lesssim 0.1$) the BAT/MAXI HR is not
  reliable; the later, low count rate, data points are shaded for
  clarity of the plot.
}
 \label{fig:HID} 
\end{figure*}

In the Hardness-Intensity diagrams (HID; Figure \ref{fig:HID}) for both BAT/MAXI 
and XRT, the grey
lines, at high count rates, show the time
during which the XRT was unable to observe, while the circles show the
start and end of the intermediate state between the soft and hard
states. Note that at low count rates ($<0.1$) the BAT/MAXI HR is not
reliable; the later, low count rate, data points are shaded for
clarity of the plot. Both HID are consistent with the canonical
trajectory of black hole binaries (Homan \& Belloni 2005). It is clear from the XRT HID that
the hardness has yet to increase to the original value, indicating
that the source has not yet returned to its original hard state,
though the count rate is lower than it was in that rising hard state,
or that the late hard state has different spectral properties (e.g. column density).

\section{Conclusions}\label{section:conclusions}

\emph{Swift} observations of the first, and so far only, observed
outburst of XTE\,J1752-223 allow us to
confirm and refine the epochs of the canonical X-ray states. 
The observations also allow us to confirm the optical counterpart 
for which we were able to produce a sub-arcsecond position.
We show that there is a correlation between optical and X-ray emission in the soft state 
as well as a  hysteresis effect where, for a given X-ray count rate, the 
magnitude in the rising hard state is significantly higher than that in the soft state.
This is similar to the hysteretical behaviour observed in the nIR for a number of transients 
though it is not normally observed in the optical bands. 
The discussed X-ray and optical, photometric, spectral and timing properties of 
XTE\,J1752-223 support its candidacy as a black hole in the Galactic centre region.

\section*{Acknowledgements}

This research has made use of
Swift data supplied by the UK Swift Science Data Centre at the University of Leicester  
and MAXI data provided by RIKEN, JAXA and MAXI teams.

\section*{References}

\re
{Brocksopp} C. et al. 2009, \atel,  2278

\re
{Curran} P.A. et al. 2010, \mnras, 410, 541

\re
{Homan} J. 2010, \atel, 2387

\re
{Homan} J.  \& {Belloni} T. 2005, \apss, 300, 107

\re
{McClintock} J.E.,  \& {Remillard} R.A. 2006, {Black hole binaries}, 157

\re
{Markwardt} C.B. et al. 2009a, \atel, 2258

\re
{Markwardt} C.B. et al. 2009b, \atel, 2261

\re
{Mu{\~n}oz-Darias} T. et al. 2010a, \atel, 2518

\re
{Nakahira} S. et al. 2009, \atel, 2259

\re
{Remillard} R.A. et al. 2009, \atel, 2265

\re
{Russell} D.M. et al. 2006, \mnras, 371, 1334

\re
{Russell} D.M. et al. 2007, \mnras, 379, 1401

\re
{Shaposhnikov} N. et al. 2009, \atel, 2269

\re
{Torres} M.A.P. et al. 2009a, \atel, 2263

\re
{Torres} M.A.P. et al. 2009b, \atel, 2268

\re
{Wilson-Hodge} C.A. et al. 2009, \atel, 2280

\re
{Yang} J. et al. 2010, \mnras, 409, L64

\newpage

\label{last}

\end{document}